\documentclass{svproc}

\usepackage{graphicx}
\usepackage[dvipsnames]{xcolor}
\definecolor{Su}{RGB}{0, 77, 154}
\definecolor{In}{RGB}{154, 0, 77}
\definecolor{Re}{RGB}{85,168,104}
\usepackage{appendix}
\usepackage[utf8]{inputenc}
\usepackage{amsmath}
\usepackage{amsfonts}
\usepackage{graphicx}
\usepackage{color}
\usepackage{subfig}
\usepackage{tikz}
\usetikzlibrary{positioning,calc}
\usepackage{color}
\usepackage[misc]{ifsym} 
\usepackage{mathtools}
\usepackage{dirtytalk}
\usepackage{amssymb}
\usepackage{bbm, dsfont}
\usepackage{comment}
\usepackage[labelfont=bf]{caption}
\usetikzlibrary{shapes,snakes}

\newcommand{\nmGA}{\texttt{nMGA}}
\newcommand{\lga}{\texttt{LGA}}

\newcommand{\redsim}{\texttt{RED-Sim}}
\newcommand{\suml}[1]{\sum\limits_{#1}}

\usepackage{url}

\begin{document}
\title{Rejection-Based Simulation of Non-Markovian Agents on Complex Networks}
\titlerunning{Rejection-Based Simulation of Non-Markovian Agents}

%
\author{Gerrit Großmann\textsuperscript{(\Letter)}\inst{1} \and \\
Luca Bortolussi\inst{1,2} \and
Verena Wolf\inst{1}}
\authorrunning{G. Großmann et al.}
%
\institute{Saarland University, 66123 Saarbrücken, Germany
\email{\{gerrit.grossmann,verena.wolf\}@uni-saarland.de}
\and
University of Trieste, Trieste, Italy \\
\email{luca@dmi.units.it}
}
\maketitle              
\begin{abstract}
Stochastic models in which agents interact with their neighborhood according to a network topology are a powerful modeling framework to study the emergence of complex dynamic patterns in real-world systems.
Stochastic simulations are often the preferred---sometimes the only feasible---way to investigate such systems.
Previous research focused primarily on Markovian models where the random time until an interaction happens follows an exponential distribution. 

In this work, we study a general framework to model systems where each agent is in one of several states. Agents can change their state at random, influenced by their complete neighborhood, while the time to the next event can follow an arbitrary probability distribution. 
Classically, these simulations are hindered by high computational costs of updating the rates of interconnected agents and sampling the random residence times from arbitrary distributions. 

We propose a rejection-based, event-driven simulation algorithm to overcome these limitations. 
Our method over-approximates the instantaneous rates corresponding to inter-event times while
rejection events counter-balance these over-approximations.
We demonstrate the effectiveness of our approach on models of epidemic and information spreading. 
\keywords{Gillespie Simulation, Complex Networks, Epidemic Modeling, Rejection Sampling, Multi-Agent System}
\end{abstract}

\section{Introduction}
Computational modeling of dynamic processes on complex networks is a thriving research area \cite{barabasi2016network,goutsias2013markovian,pastor2015epidemic}.
Arguably, the most common formalism for spreading processes is the continuous-time $\texttt{SIS}$ model and its variants \cite{kiss2016mathematics,porter2016dynamical,rodrigues2016application}.
Generally speaking, an underlying contact network specifies the connectivity between nodes (i.e., agents) and each agent occupies one of several mutually exclusive (local) states (or compartments). In the well-known $\texttt{SIS}$ model, these states are \emph{susceptible} ($\texttt{S}$) and \emph{infected} ($\texttt{I}$). Infected nodes can recover (become susceptible again) and propagate their infection to neighboring susceptible nodes. 

$\texttt{SIS}$-type models have shown to be extremely useful for analyzing and predicting the spread of opinions, rumors, and memes in online social networks \cite{kitsak2010identification,zhao2012sihr} as well as the neural activity \cite{goltsev2010stochastic,meier2017epidemic}, the spread of malware \cite{gan2012propagation}, and blackouts in financial institutions \cite{may2009systemic,peckham2013contagion}.

Previous research focused mostly on models where the probability of an event (e.g. infection or recovery) happening in the next (infinitesimal) time unit is constant, i.e. independent of the time the agent has already spent in its current state. 
We call such agents \emph{memoryless} and the corresponding stochastic process \emph{Markovian}. 
The semantics of such a model can be described using a so-called (discrete-state) \emph{continuous-time Markov chain} (CTMC). 

One particularly important consequence of the memoryless property is that the random time until an agent changes its state, either because of an interaction with another agent or because of a spontaneous transition, follows an exponential distribution. 
The distribution of this \emph{residence time} is parameterized by an (interaction-specific) rate $\lambda \in \mathbb{R}_{\geq 0}$ \cite{kiss2016mathematics}. 
Each agent has an associated event-modulated Poisson process whose rate depends on the agent's state and the state of its neighbors \cite{masuda2018gillespie}. 
For instance, infection of  an agent increases the rate at which its susceptible neighbors switch to the infected state. 

However, exponentially distributed residence times are an unrealistic assumption in many real-word systems.
This holds in particular for the spread of epidemics \cite{lloyd2001realistic,yang1972empirical,blythe1988variable,hollingsworth2008hiv,feng2000endemic}, for the diffusion of opinions in online social networks \cite{barabasi2005origin,vazquez2006modeling}, and interspike times of neurons \cite{softky1993highly} as empirical results show. 
However, assuming time delays that can follow non-exponential distributions complicate the analysis of such processes and typically only allow Monte-Carlo simulations, which suffer from high  computational costs.

Recently, the Laplace-Gillespie algorithm (\lga{})  for the simulation of non-Markovian dynamics has been introduced by Masuda and Rocha in \cite{masuda2018gillespie}. 
It is based on the non-Markovian Gillespie algorithm by Bogun{\'a} et al\;(\nmGA{}) \cite{boguna2014simulating} and minimizes the costs of sampling inter-event times.
However, both methods require computationally expensive updating of an agent's neighborhood in each simulation step, which renders them inefficient for large-scale networks. 
In the context of Markovian processes on networks, it has recently been shown that rejection-based simulation can overcome this limitation \cite{cota2017optimized,st2018efficient,grossmann2019rejection}.

Here, we extend the idea of rejection-based simulation  to  non-Markovian networked systems, 
proposing \redsim{}, a \emph{r}ejection-based, \emph{e}vent-\emph{d}riven \emph{sim}ulation approach.
Specifically, we combine three ideas to obtain an efficient simulation of non-Markovian processes: (i) we express the distributions of inter-event times through time-varying instantaneous rates, (ii) we sample events based on an over-approximation of these rates and compensate via a rejection step, and (iii) we use a priority queue to sample the next event. 
The combination of these elements makes it possible to reduce the time-complexity of each simulation step. Specifically, if an agent changes its state, no update of the rate of neighboring agents is necessary. 
This comes at the costs of rejection events to counter-balance missing information about the neighborhood.
However, using a priority queue renders the computational burden of each rejection event very small.

The remainder of the paper is organized as follows: We describe our framework for non-Markovian dynamics in Section \ref{model} and provide a review of previous simulation approaches in Section \ref{relatedwork}. Next, we propose our rejection-based simulation algorithm in Section \ref{approach}.
Section \ref{casestudies} presents numerical results and we conclude our work in Section \ref{conclusion}.

\section{Multi-Agent Model\label{model}}
This section introduces the underlying formalism to express agent-based dynamics on networks.
Let $\mathcal{G}=(\mathcal{N}, \mathcal{E})$ be a an undirected,  finite graph without self-loops, called \emph{contact network}. Nodes $n\in\mathcal{N}$ are also referred to as \emph{agents}. 
\subsubsection{Network State.} 
The current state of a network $\mathcal{G}$ is described by two functions:
\begin{itemize}
    \item $S: \mathcal{N} \rightarrow \mathcal{S}$ assigns to each agent $n$ a local state $S(n)\in \mathcal{S}$, where $\mathcal{S}$ is a finite set of local states (e.g., $\mathcal{S}=\{\texttt{S}, \texttt{I}\}$ for the $\texttt{SIS}$ model);
    \item $R: \mathcal{N} \rightarrow \mathbb{R}_{\geq 0}$, describes the residence time of each agent,  i.e. the time since the last change of state of the agent.
\end{itemize}
We say that an agent \emph{fires} when it changes its state and refer to the remaining time until it fires as its \emph{time delay}. 
The \emph{neighborhood state} $M(n)$ of an agent $n$ is a multi-set
containing the states of all neighboring agents together with their respective residence times:
\begin{equation*}
    M(n) = \Big\{ \big(S(n'),R(n')\big)  \; \big| \;  (n,n') \in \mathcal{E} \Big\} \;.
\end{equation*}
The set of all possible neighborhood-states of all agents in a given system is denoted by $\mathcal{M}$.

\subsubsection{Network Dynamics.} The dynamics of the network is described by assigning to each agent $n$ two functions $\phi_n$ and $\psi_n$:
\begin{itemize}
    \item $\phi_n: \mathcal{S} \times \mathbb{R}_{\geq 0} \times \mathcal{M} \rightarrow  \mathbb{R}_{\geq 0}$ defines the \emph{instantaneous rate} of $n$, i.e. if $\lambda =\phi_n \big(S(n),R(n),M(n) \big)$, then the probability that $n$ fires in the next infinitesimal time interval $t_{\Delta}$ is  $\lambda t_{\Delta}$;  
    \item $\psi_n: \mathcal{S} \times \mathbb{R}_{\geq 0} \times \mathcal{M} \rightarrow  P_{\mathcal{S}}$ determines the   state probabilities when a transition occurs. Here,  $P_{\mathcal{S}}$ denotes the set of all probability distributions over $\mathcal{S}$.  
    Hence if $p=\psi_n \big(S(n),R(n),M(n) \big)$, then, when agent $n$ fires, the next local state is $s$ with probability $p(s)$.
\end{itemize}
Note that we do not consider   cases of pathological behavior here, e.g. where $\phi_n$ is defined in such a way that an infinite amount of simulation steps is possible in  finite time.  

A multi-agent network model is completely specified by a tuple $(\mathcal{G},$ $\mathcal{S},$ $\{\phi_n\},$ $\{\psi_n\},S_0)$, where $S_0$ denotes a function that assigns to each node an initial state.

\subsubsection{Example.\label{sisexample}}
In the classical  \texttt{SIS} model we have $\mathcal{S}=\{\texttt{S},\texttt{I}\}$ and $\phi$ and $\psi$ are the same for all agents, 
i.e., 
\begin{equation*}
   \phi_n(s,t,m)= 
    \begin{cases}
     c_r  \text{\hspace{2.15cm} if $s=\texttt{I}$} \\
     c_i  \suml{s',t' \in m} \mathbbm{1}_{\texttt{I}}(s') \text{ \; if $s=\texttt{S}$} 
    \end{cases}
   \hspace{0.1cm} \psi_n(s,t,m)= 
    \begin{cases}
        \mathbbm{1}_{\texttt{S}}  \text{\hspace{0.3cm} if $s=\texttt{I}$} \\
        \mathbbm{1}_{\texttt{I}}   \text{\hspace{0.3cm} if $s=\texttt{S}$} 
    \end{cases}
\end{equation*}
Here, $c_i, c_r \in \mathbb{R}_{\geq 0}$ denote the infection  and recovery rate constants, respectively. Note that the infection rate is proportional to the number of infected neighbors whereas the rate of recovery is independent from  neighboring agents. 
Moreover, $\mathbbm{1}_{s}: \mathcal{S} \rightarrow \{0,1\} $ maps a state $s'$ to one iff $s=s'$ and to zero otherwise. 
The model is Markovian as neither $\phi$ nor $\psi$ depend on the residence time of any agent.
\subsection{Semantics\label{semantics}}
We will specify the semantics of a multi-agent model by describing a   stochastic simulation algorithm that generates  trajectories of the system.
It is  based on a race condition among agents:  each agent picks a random time until it will fire, but only the one with the shortest time delay wins and changes its state.

\subsubsection{Time Delay Density.} 
Assume that $t_{\Delta}$ is the time increment of the algorithm. 
We  define for each $n$ the \emph{effective rate} $\lambda_n(t_{\Delta})$ as
$$
\begin{array}{l}
   \lambda_{n}(t_{\Delta}) = \phi_n\Big(S(n),R(n)+t_{\Delta}, M_{t_{\Delta}}(n) \Big),\mbox{ where}
   \\[1ex]
M_{t_{\Delta}}(n) = \Big\{ \big(S(n'),R(n')+t_{\Delta}\big)  \; \big| \;  n,n' \in \mathcal{E} \Big\}, 
\end{array}
$$
describes the neighborhood-state of $n$ in $t_{\Delta}$ time units assuming that all agents remain in their current state. 
Next we assume that for each node $n$,   the probability density   of the (non-negative) time delay  is $\gamma_n$, i.e. 
$\gamma_n(t_{\Delta})$ is the density of firing after $t_{\Delta}$ time units.
Leveraging the theory of renewal processes \cite{cox1962renewal},
we find the relationship 
\begin{equation}\label{eq:ratefunc}
    \lambda_n(t_{\Delta}) =\frac{\gamma_n(t_{\Delta})}{1-\int_{0}^{t_{\Delta}} \gamma_n(t_{\Delta})}
    \hspace{0.7cm} \text{and} \hspace{0.7cm} 
    \gamma_n(t_{\Delta}) = \lambda_n(t_{\Delta}) e^{- \int_0^{t_{\Delta}} \lambda_n(y)dy}
    \;.
\end{equation}
We assume $ \lambda_n(t_{\Delta})$ to be zero if the denominator is zero. 
Note that using this equation, we can  derive rate functions from a given time delay distribution (i.e. uniform, log-normal, gamma, and so on). 
If it is not possible to derive $\lambda_n$ analytically, it can be computed numerically. 

\begin{figure}
  \centering
    \subfloat[]{\includegraphics[width=0.245\textwidth]{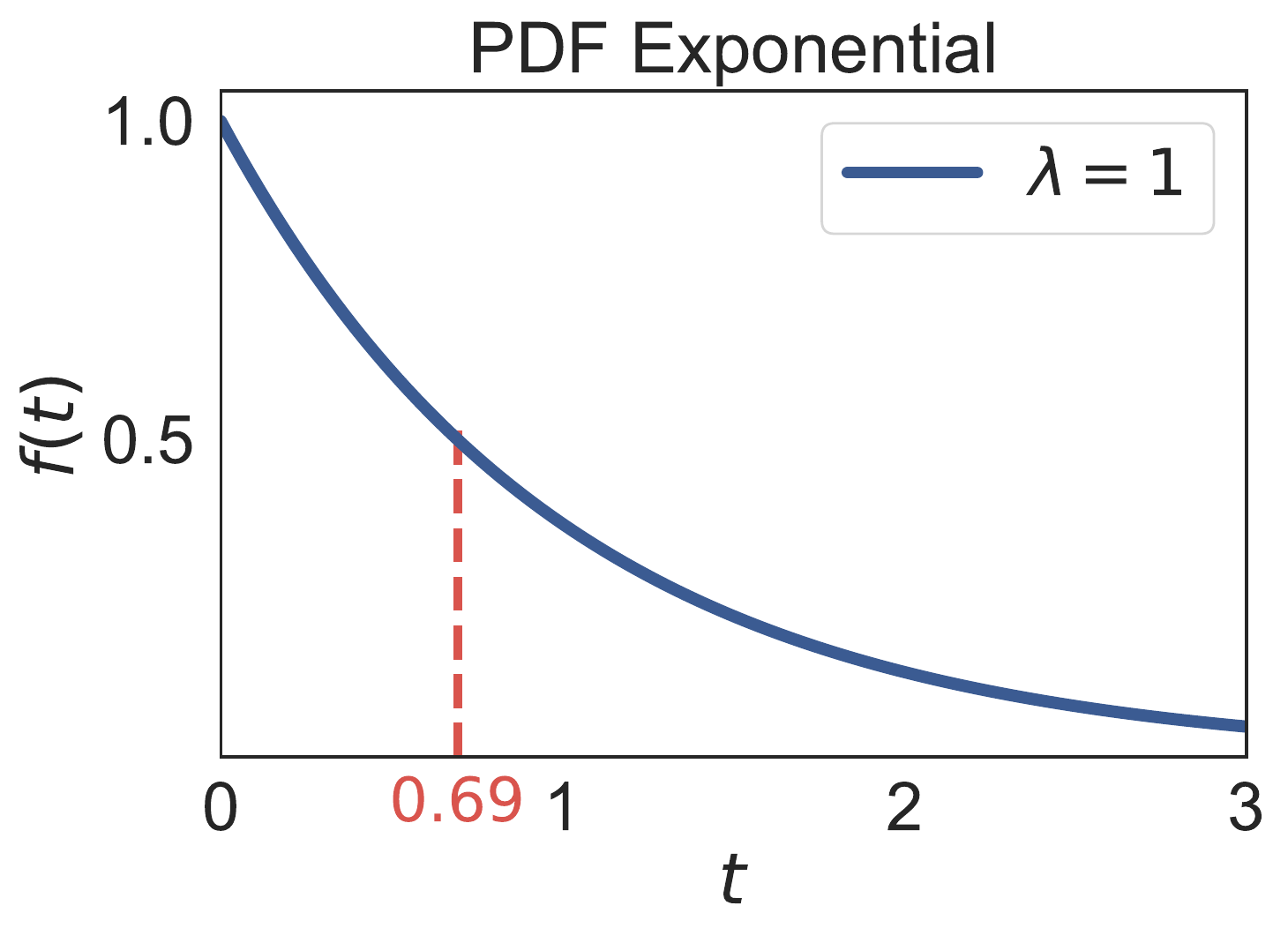}}
    \subfloat[]{\includegraphics[width=0.245\textwidth]{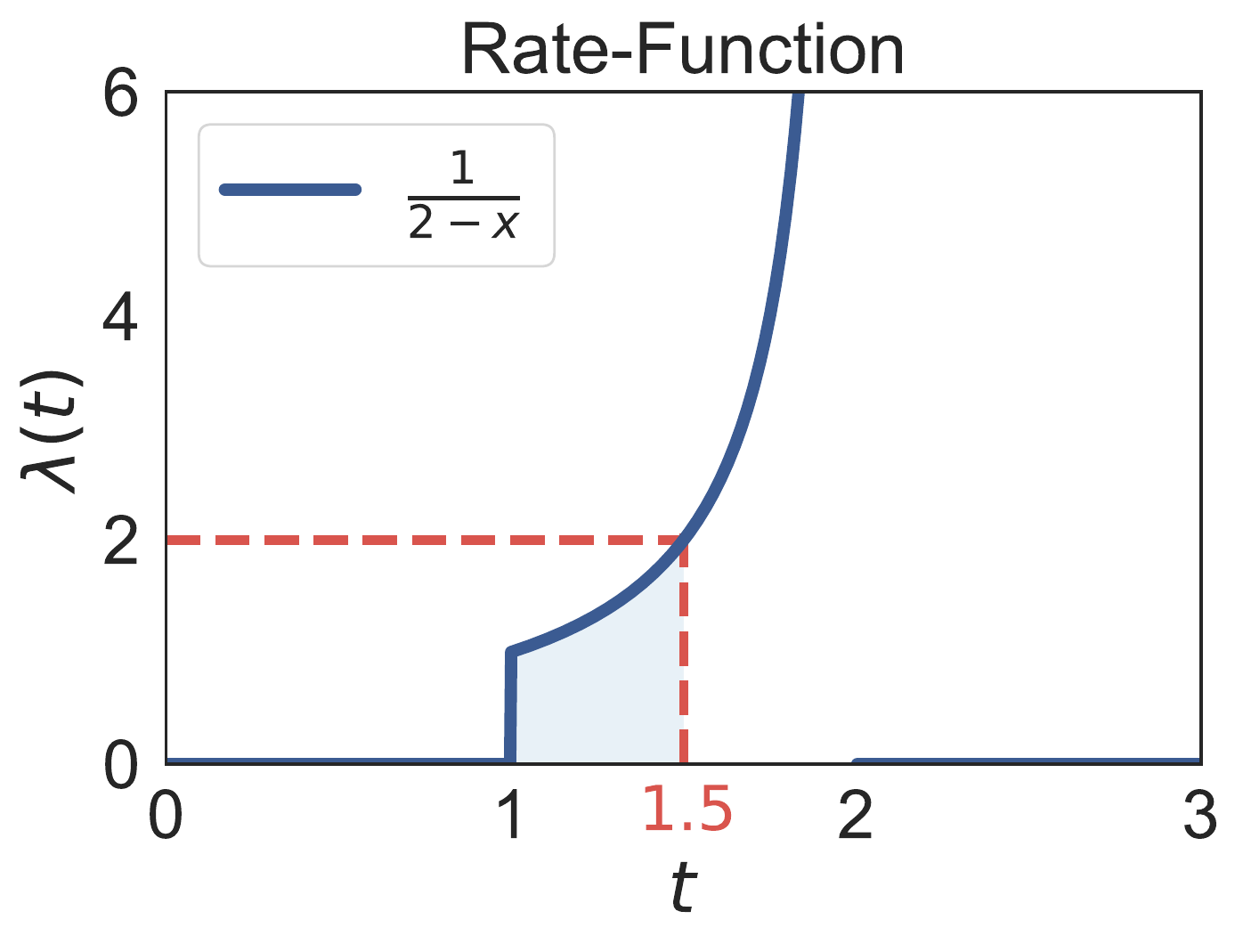}} 
    \subfloat[]{\includegraphics[width=0.245\textwidth]{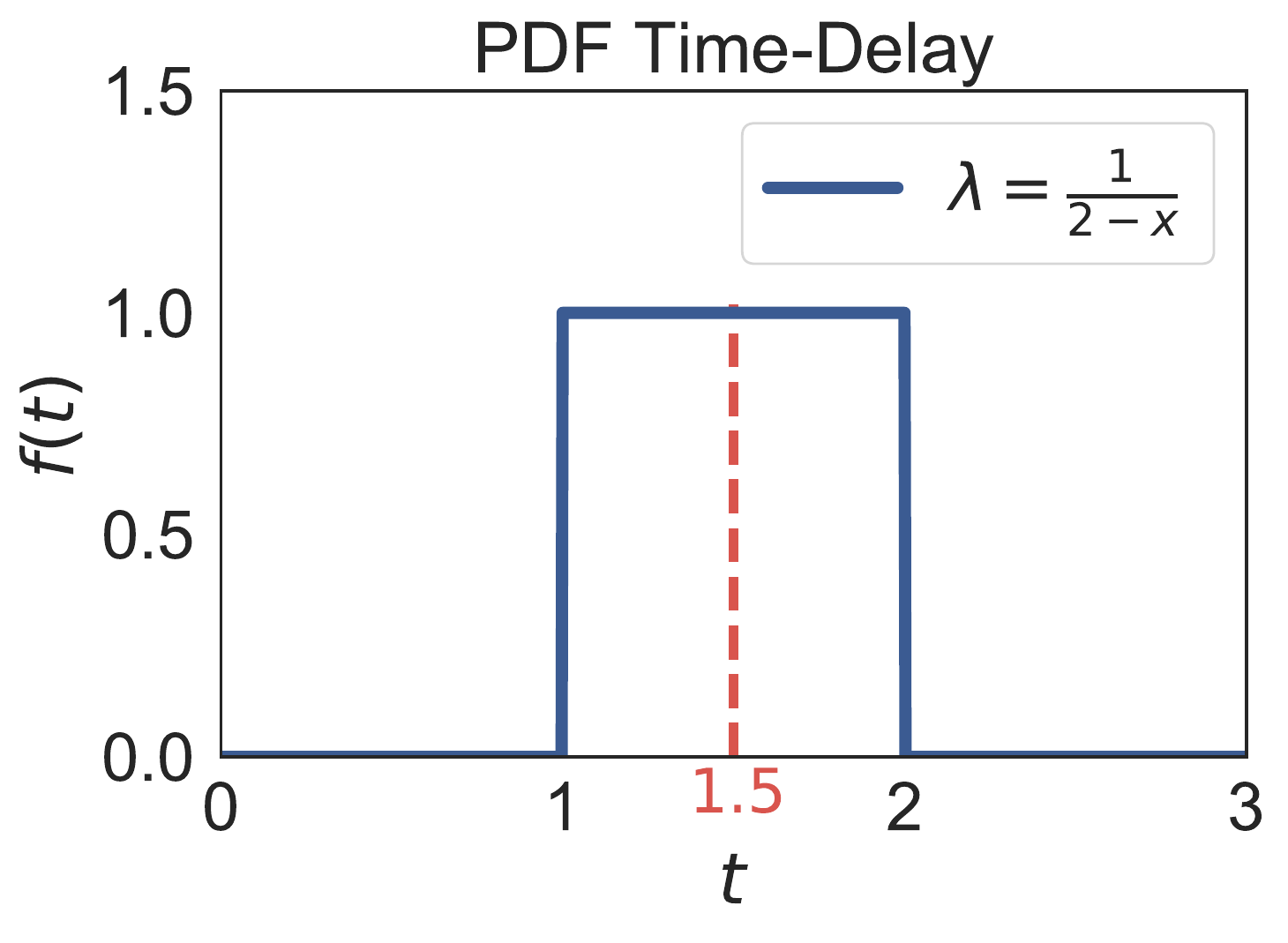}}
    \subfloat[]{\includegraphics[width=0.245\textwidth]{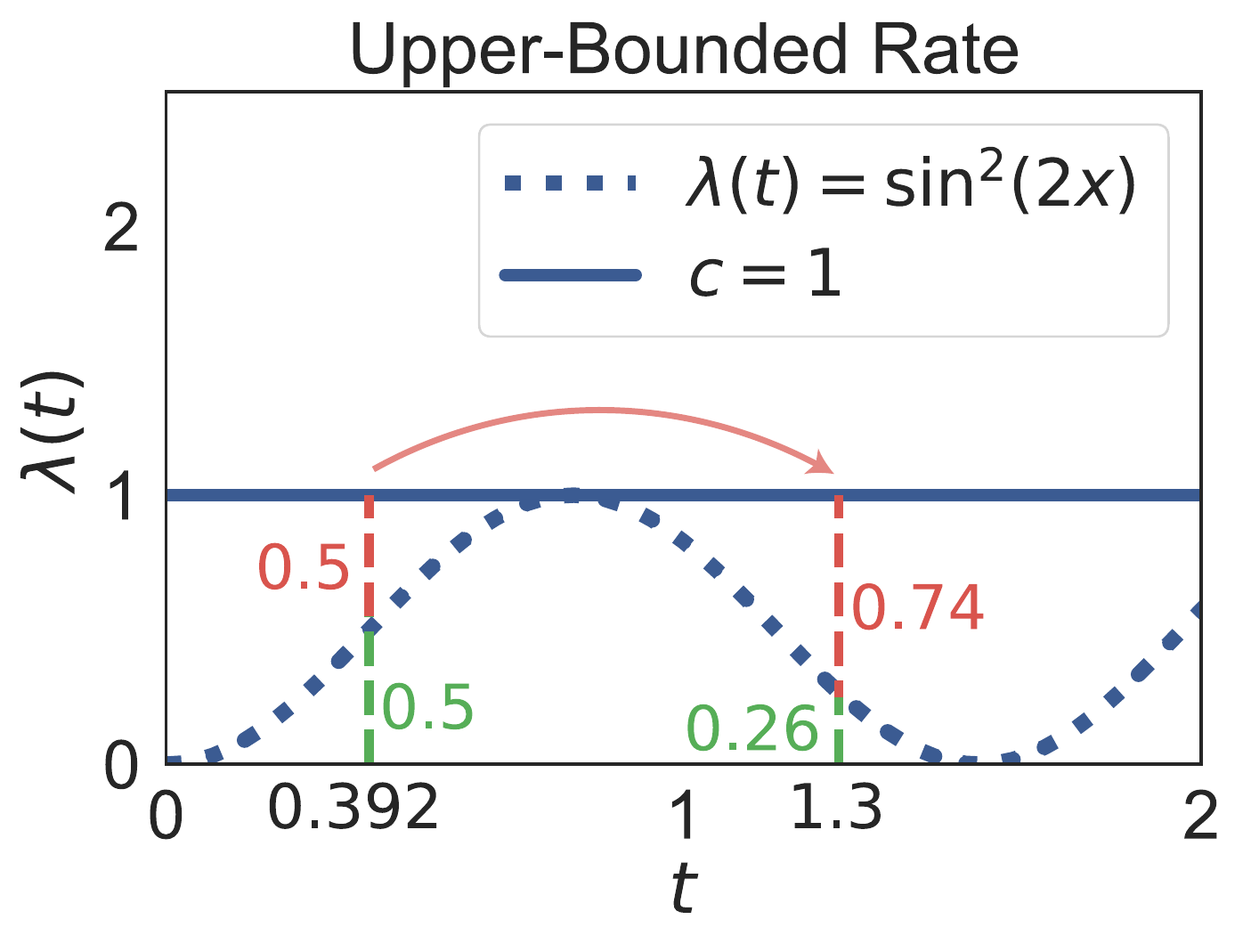}}
      \caption{(a-c) Sampling event times with a rate function $\frac{\mathds{1}_{t \in [1,2] }}{2-t}$. (a) Generate a random variate from the exponential distribution with rate $\lambda=1$, the sample here is $0.69$. (b) We integrate the rate function until the area is $0.69$, here $t_n = 1.5$. (c) This is the rate function corresponding to the uniform distribution in $\gamma(t)=\mathds{1}_{t \in [1,2] }$. (d) Sampling $t_n$ from a time-varying rate function using an upper-bound of $c=1$, rejection probabilities shown in red.
      }
     \label{fig:genPDF}
\end{figure}

For example, a constant rate function $\lambda(t_{\Delta})=c$ corresponds to an exponential time delay distribution $\gamma(t_{\Delta})=ce^{-ct_{\Delta}}$ with rate $c$. 
Fig.~\ref{fig:genPDF} (b) illustrates   the rate function when $\gamma$ is the uniform distribution on $[1,2]$.

\subsubsection{Sampling Time Delays.\label{sample}}
The effective rate $\lambda_{n}(t_{\Delta})$ allows us to sample the time delay $t_n$ after which agent $n$ fires, using the inversion method. 
First, we sample an exponential random variate $x$ with rate $1$,  then we integrate $\lambda_{n}(t_{\Delta})$ to find $t_n$ such that  
\begin{equation}
    \int_0^{t_n} \lambda_n(t_{\Delta}) dt_{\Delta} =x \;. \label{eq:integralsampl}
\end{equation}
In general it is possible to pre-compute the integral \cite{pasupathy2010generating}, but its parameterization (on states, residence times, etc) renders this difficult.

Another viable approach is to use rejection sampling. Assume that we have  $c\in \mathbb R_{\ge 0}$   such that $\lambda_n(t_{\Delta}) \leq c$ for all $t_{\Delta}$. 
We start with $t_n=0$. In each step, we sample an exponentially distributed random variate $t_n'$ with rate $c$ and set $t_n = t_n+t_n'$. We accept $t_n$ with probability $\frac{\lambda_n(t_n)}{c}$.
Otherwise we reject it and repeat the process. 
If a reasonable over-approximation can be constructed, this is typically much faster than the integral approach in \eqref{eq:integralsampl}.

\subsubsection{Na\"ive Simulation Algorithm.\label{naive}}
The following simulation algorithm generates statistically correct trajectories of the model.
 It starts by initializing the global clock $t_{global}=0$ and setting $R(n)=0$ for all $n$. 
The algorithm repeatedly performs simulation steps until a predefined time horizon or some other stopping criterion is reached. Each stimulation step is as follows:
\begin{enumerate}
\item Generate a random time delay (candidate) $t_n$ for each agent $n$ using $\gamma_n$. Identify agent $n'$ with the smallest time delay $t_{n'}$.
\item Pick the next state $s'$ for $n'$ according to $\psi_{n'} \big(S(n'),R(n')+t_{n'}, M(n')\big)$ and set $S(n')=s'$. Set $R(n')=0$ and 
$R(n)=R(n)+t_{n'}$ ($\forall n \neq n'$).
\item Set $t_{global}=t_{global}+t_{n'}$ to update the global clock and go to Line \texttt{1}. 
\end{enumerate}
Note that this algorithm is very inefficient as it requires an expensive iteration over all agents and sampling of time delays in each  step.

\section{Previous Simulation Approaches \label{relatedwork}}
Most recent work on non-Markovian dynamics focuses on the mathematical modeling of such processes \cite{kiss2015generalization,pellis2015exact,jo2014analytically,sherborne2016mean,starnini2017equivalence}. 
In particular, research has focused on how specific distributions (e.g. constant recovery times) alter the properties of epidemic spreading such as the epidemic threshold (see \cite{pastor2015epidemic,kiss2016mathematics} for an overview). 
However, only few approaches are known for the simulation of non-Markovian dynamics \cite{boguna2014simulating,masuda2018gillespie}.
We shortly review them in the sequel.

\subsection{Non-Markovian Gillespie Algorithm}
Bogun{\'a} et al.\;propose a direct generalization of the Gillespie algorithm to non-Markovian systems, \nmGA{}, which is statistically exact but computationally expensive \cite{boguna2014simulating}. 
The algorithm is conceptually similar to our baseline in Section \ref{semantics} but computes the time delay using so-called \emph{survival functions}. 
An agent's survival function determines the probability that its time delay is larger than a certain time $t_{\Delta}$. The joint survival function of all agents determines the probability that all time delays are larger than $t_{\Delta}$ which can be used to sample the next event time.

The drawback of the \nmGA{} is that it is necessary to iterate over all agents in each step in order to construct their joint survival function. As a fast approximation, the authors suggest to only use the current instantaneous rate at $t=0$ \big(i.e., $\lambda_{n}(0)$\big) and assume all rates remain constant until the next event. This is correct in the limit of infinite agents, because when the number of agent approaches infinity,  the time until the next firing of any agent approaches zero. 

\subsection{Laplace-Gillespie Algorithm}
The \lga{}, introduced by Masuda and Rocha in \cite{masuda2018gillespie}, aims at reducing the computational cost of finding the next event time compared to $\nmGA$, while remaining statistically correct. 
It assumes that the time delay distributions can be expressed in the form of a weighted average of exponential distributions
\begin{equation*}
    \gamma_n(t_{\Delta}) = \int_{0}^{\infty} p_n(\lambda)\lambda e^{-\lambda t_{\Delta}} d \lambda \;,
\end{equation*}
where $p_n$ is a PDF over the rate $\lambda \in \mathbb{R}_{\geq 0}$.  
This formulation of $\gamma_n$, while being very elegant, limits the applicability to cases where the corresponding survival function is \emph{completely monotone} \cite{masuda2018gillespie}, which limits the set of possible inter-event time distributions. 

The \lga{} has two advantages. Firstly, we can sample $t_n$ by first sampling $\lambda$ according to $p_n$ and then, instead of the numerical integration in Eq.\;\eqref{eq:integralsampl},
compute $t_n=- \ln u/\lambda$ where $u$ is uniformly distributed on $(0,1)$.
Secondly, we can assume that the sampled $\lambda$ for a particular agent remains constant until one of its neighbors fires. Thus, in each step, it is only necessary to update the rates of the neighbors of the firing agent, and not of all agents.

\section{Our Method\label{approach}}
Rejection sampling for the efficient simulation of Markovian stochastic processes on complex networks has been proposed recently \cite{cota2017optimized,st2018efficient,grossmann2019rejection,vestergaard2015temporal}, but not for   the non-Markovian case where arbitrary distributions for the inter-event times are considered.

Here, we proposes the \redsim{} algorithm for the generation of statistically correct simulations of non-Markovian network models, as described in Section \ref{model}. 
The main idea of \redsim{} is to rely on rejection sampling to reduce the computational cost, making it unnecessary to update the rates of the neighbors of a firing agent. Independently from that, rejection sampling can also be utilized to sample $t_n$ without numerical integration. 

\subsection{Rate Over-Approximation}
Recall that  $\lambda_n(\cdot)$ expresses how the instantaneous rate of $n$ changes over time, assuming that no neighboring agent changes its state. 
A key in ingredient of our method is now $\widehat{\lambda}_n(\cdot)$ which upper-bounds the instantaneous rate of $n$, assuming that all neighbors are allowed to freely change their state as often as possible. That is, at all times $\widehat{\lambda}_n(t_{\Delta})$ is an upper-bound of $\lambda_n(t_{\Delta})$ taking into consideration all possible states of  the neighborhood. 

Consider again the Markovian \texttt{SIS} example. The curing of an infected node does not depend on an agent's neighborhood anyway. The rate is always $c_r$, which is a trivial upper bound. A susceptible node becomes infected with rate $c_i$ times \say{number of infected neighbors}. Thus, the instantaneous infection rate of an agent $n$ can be bounded by  $ \widehat{\lambda}_n(t_{\Delta})=k_n c_i$ where $k_n$ is the degree of $n$. 
Upper-bounds may also be constant  or depend on time.
Consider for example a  recovery time that is uniformly distributed on $[1,2]$. 
In this case,  $\lambda_n(\cdot)$ approaches infinity (cf.~Fig.~\ref{fig:genPDF}b) making a constant upper-bound impossible. 
For multi-agent models, a time-dependent upper-bound always exists since we can compute the maximal instantaneous rate w.r.t. all reachable  neighborhood states.

\subsection{The \redsim{} Algorithm \label{algo}}
For a given multi-agent model specification $(\mathcal{G},\mathcal{S},\{\phi_n\},\{\psi_n\},S_0)$ and given upper-bounds $\{\widehat{\lambda}_n\}$, we propose a statistically exact simulation algorithm, which is based on two basic data structures: 
\begin{description}
\item[Labeled Graph]\hfill \\
A graph represents the contact network and each agent (node) $n$ is annotated with its current state $S(n)$ and $T(n)$, the time point of its last state change. 
\item[Event Queue]\hfill \\
The event queue stores the list of future events, where an event is a tuple $(n,\widehat{\mu},\widehat{t}_n)$. 
Here, $n$ is the agent that fires, $\widehat{t}_n$ the prospective absolute time point of firing, and $\widehat{\mu} \in \mathbb{R}_{\geq 0}$ 
is an over-approximation of the true effective rate (at time point $\widehat{t}_n$).
The queue is sorted w.r.t.\;$\widehat{t}_n$ and initialized by generating one event per agent.
\end{description}
A global clock, $t_{\text{global}}$, keeps track of the elapsed time since the simulation started.  
We use $T(n)$ instead of $R(n)$ to avoid updates for all agents after each event (i.e., $R(n)=t_{\text{global}}-T(n)$).
We perform simulation steps until some termination criterion is fulfilled, each step is as follows:
\begin{enumerate}
\item Take the first event $(n,\widehat{\mu},\widehat{t}_n)$ from the event queue and update $t_{\text{global}}=\widehat{t}_n$.
\item Evaluate the true instantaneous rate  $\mu = \phi_n \big(S(n),t_{\text{global}}-T(n),M(n)\big)$ of $n$ at the current system state.
\item With probability $1-\frac{\mu}{\widehat{\mu}}$, \textbf{reject} the firing  and go to Line \texttt{5}. 
\item Randomly choose the next state $s'$ of $n$ according to the distribution \\ $\psi_n \big(S(n),t_{\text{global}}-T(n), M(n)\big)$. If $S(n)\neq s'$: set $S(n)=s'$ and $T(n)=t_{\text{global}}$.
\item Generate a new event for agent $n$ and push it to the event queue. 
\item Go to Line \texttt{1}. 
\end{enumerate}

The correctness of \redsim{} can be shown similarly to \cite{grossmann2019rejection,cota2017optimized} (see also the proof sketch in Appendix \ref{app:correctness}).
Note that  in all approaches 
evaluating an agent's instantaneous rate is linear in the number of its neighbors. In previous approaches, the rate has to be updated for all neighbors of a firing agent. In \redsim{} only the rate of the firing agent has to be updated. 
The key asset of \redsim{} is that, due to the over-approximation of the rate function, we do not need to update the neighborhood of the firing agent $n$, even though the neighbor's respective rates might change as a result from the event.
We   provide a more detailed analysis of the time-complexity of \redsim{}  in Appendix \ref{app:runtime}. 

\vspace{-0.05cm}
\subsubsection{Event Generation.}
To generate new events in Line \texttt{5}, we sample a random time delay $t_n$ and set
$\widehat{t}_n= t_{\text{global}} + t_n$. To sample $t_n$
according to the over-approximated rate  $\widehat{\lambda}_n(\cdot)$,
we either use the integration approach of Eq.\;\eqref{eq:integralsampl} or sample directly from an upper-bounded the exponential distribution (cf.\;Fig.\;\ref{fig:genPDF}d).

To sample $t_n$ from an exponential distribution, we need to be able to find an upper bound that is constant in time $\widehat{\lambda}_n(t)=c$ for all $t$. 
Hence, we simply set $\widehat{\mu}=c$ and sample $t_n$ from an exponential distribution with rate $c$.
Otherwise, when a constant upper bound either does not exits or is unfeasible to construct, we use numerical integration over $\widehat{\lambda}_n(\cdot)$ (see Eq. \eqref{eq:integralsampl}), and set $\widehat{\mu}=\widehat{\lambda}_n(t_n)$.  
Alternatively, when $\widehat{\lambda}_n(t)$ has the required form (cf.\;Section \ref{relatedwork}), we can even use \lga{}-like approach to sample $t_n$ \cite{boguna2014simulating}
(and also set $\widehat{\mu}=\widehat{\lambda}_n(t_n)$).

\subsubsection{Discussion.}
We expect \redsim{}  to perform poor   only in some   special cases, where either the construction of an upper-bound is numerically too expensive or where the difference between the upper-bound and the actual average rate is very large, which would render the number of rejections events too high.

It is easy to extend \redsim{} to different types of non-Markovian behavior. For instance, we might keep track of the number of firings of an agent and parameterize $\phi$ and $\psi$ accordingly to generate the behavior of self-exiting point processes or to cause correlated firings among agents \cite{ogata1981lewis,dassios2013exact}. 

Note that, we can turn the method into a rejection-free approach by generating a new event for $n$ and all of its neighbors in Line \texttt{5} while taking the new state of $n$ into consideration (see also Appendix \ref{app:correctness}). 

\section{Case Studies\label{casestudies}}
\begin{figure}[t!]
  \centering
  \hspace{0.1cm}
    \subfloat[]{\includegraphics[width=0.48\textwidth]{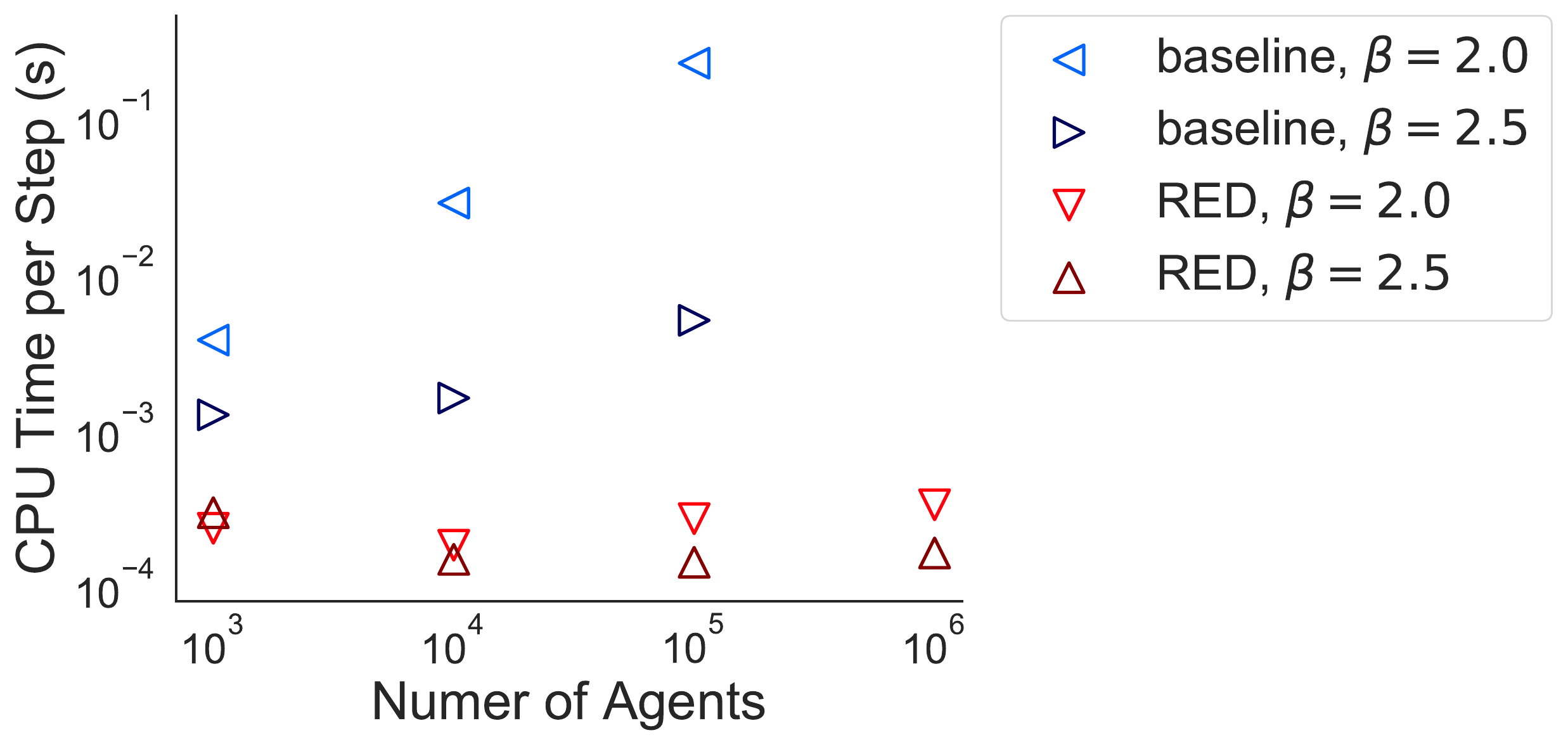}}
    \hspace{0.01cm}
    \subfloat[]{\includegraphics[width=0.48\textwidth]{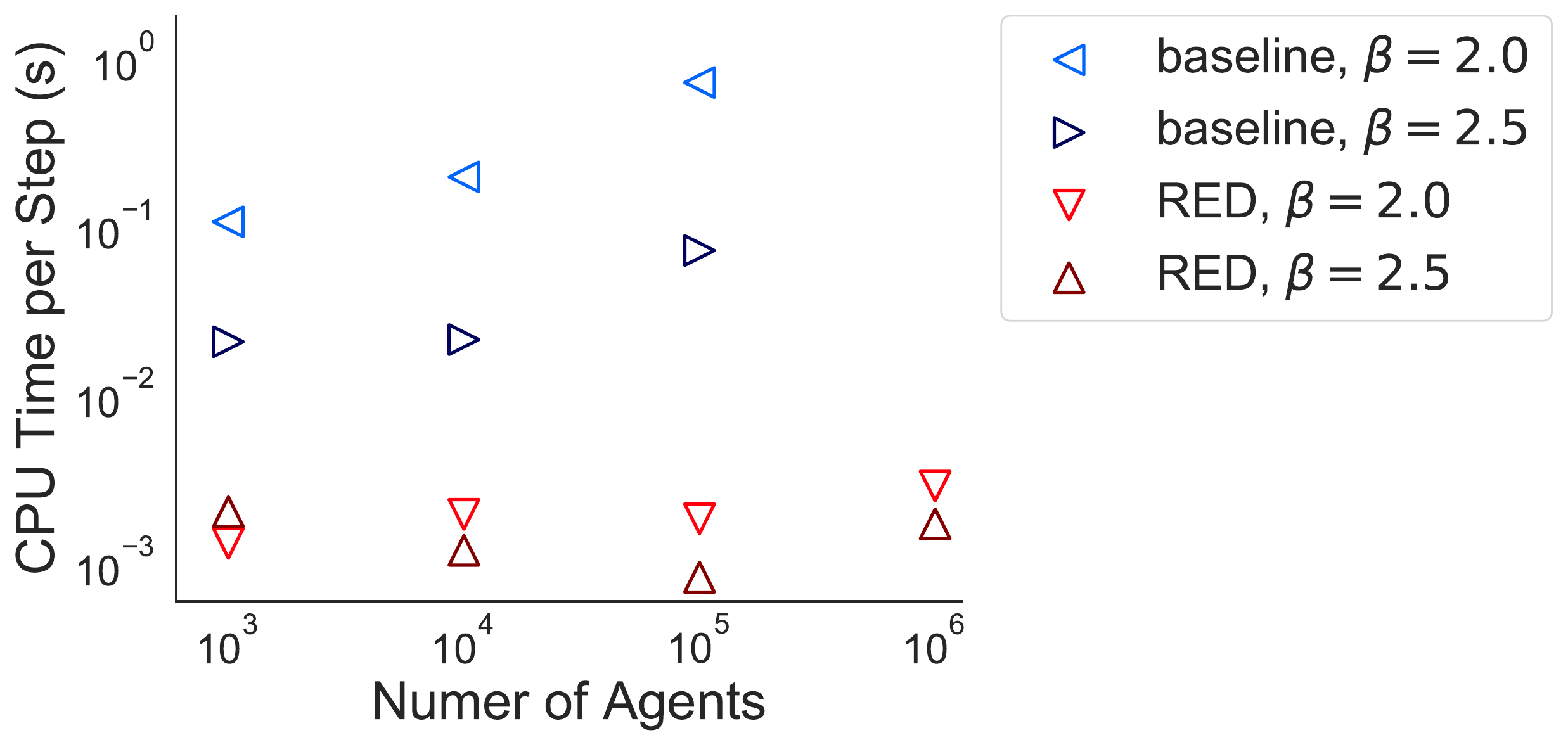}}
      \caption{Computation time of a single simulation step w.r.t.~network size and connectivity of the \textbf{{SIS} model (a)} and \textbf{voter model (b)}. We measure the CPU time per simulating step by dividing the simulation time by the number of successful (i.e., non-rejection) steps.}
     \label{fig:Results}
\end{figure}

We demonstrate the effectiveness our approach on classical epidemic-type processes and synthetically generated networks following the configuration model with a truncated power-law 
degree distribution \cite{fosdick2018configuring}. 
That is $P(k) \propto k^{-\beta}$ for $3 \leq k \leq |\mathcal{N}|$. We use $\beta \in \{2,2.5\}$ (a smaller $\beta$ corresponds to a larger average degree). 
The implementation is written in Julia and publicly  available\footnote{\texttt{github.com/gerritgr/non-markovian-simulation}}. As a baseline for comparison, we use the rejection-free variant of the algorithm where neighbors are updated after an event (as described at the end of Section \ref{algo}).
The evaluation was performed on a 2017 MacBook Pro with a 3.1 GHz Intel Core i5 CPU and results are shown in Fig.~\ref{fig:Results}.
\vspace{-0.3cm}
\subsubsection{\texttt{SIS} Model.}
We consider an \texttt{SIS} model (with $\psi$ and $\phi$ as defined above), but infected nodes become less infectious over time. That is, the rate at which an infected agent with residence time $t$ \say{attacks} its susceptible neighbors is $u e^{- u t}$ for $u=0.4$. This shifts the exponential distribution to the left. 
We upper-bound the infection rate of an agent $n$ with degree $k_n$ with $\widehat{\lambda}_n(t)=u k_n$ which is constant in time. Thus, we sample $t_n$ using an exponential distribution. 
The time until an infected agent recovers is, independent from its neighborhood, uniformly distributed   in $[0,1]$ (similar to \cite{rost2016impact}). 
Hence, we can sample it directly.  
We start with $5\%$ infected agents. 
\vspace{-0.3cm}
\subsubsection{Voter Model.}
The voter model describes the competition of two opinions of agents in state $\texttt{A}$ switch to $\texttt{B}$ and vice versa (i.e. $\psi$ is deterministic).
The time   until an agent switches follows a Weibull distribution (similar to \cite{boguna2014simulating,van2013non}):
\begin{equation*}
    \gamma_n(t)=c u (t u )^{c-1}  e^{-{(t u)}^{c}}
    \hspace{0.5cm} \text{and}\hspace{0.5cm}
    \lambda_n(t) = c u (t u )^{c-1}, \; t\ge 0
\end{equation*}
where we set $c=c_A=2.0$, $u=u_A$ if $S(n)=\texttt{A}$ and $c=c_B=2.05$, $u=u_B$ if $S(n)=\texttt{B}$.
We let the fraction of opposing neighbors modulate $u$, i.e., $u_A = \frac{B_n}{k_n}$, where $B_n$ denotes the number of neighbors currently in state \texttt{B} and $k_n$ is the degree of agent $n$ (and analogously for \texttt{A}).
Hence, the instantaneous rate depends on the current residence time and the states of the neighboring agents. 
To get an upper-bound for the rate, we set  $u_A=u_B=1$ and get $\widehat{\lambda}_n(t) = c t ^{c-1}$.
We use numerical integration to sample $t_n$ to show that \redsim{} performs well also in the case of this more costly sampling.  We start with $50\%$ of agents being in each state. 
\vspace{-0.3cm}
\subsubsection{Discussion.}
Our results provide strong evidence for the usefulness of rejection sampling for non-Markovian simulation. 
As expected, we find that the number of interconnections (edges) and the number of agents influence the runtime behavior. Especially for \redsim{}, the number of edges shows to be much more relevant than purely the number of agents. 
Our method consistently outperforms the baseline up to several orders of magnitude. 
The gain (\redsim{} speed by baseline speed) ranges from $ 10.2$ ($10^3$ nodes, voter model, $\beta=2.5$) to $ 674$  ($10^5$ nodes, \texttt{SIS} model, $\beta=2.0$).

We expect the baseline algorithm to be comparable with \lga{} as both of them only update the rates of the relevant agents after an event. 
Moreover, in the \texttt{SIS} model, sampling the next event times is very cheap. However, a detailed statistical comparison remains to be performed (both case-studies could not straightforwardly be simulated with \lga{} due to its constraints on the time delays).
Note that, when \lga{} is applicable, its key asset, the fast sampling of time delays, can also be used in \redsim{}.
We also tested a \texttt{nMGA}-like implementation where rates are consider to remain constant until the next event. However, the method was---even though it is only approximate---slower than the baseline. 

Note that the \texttt{SIS} model is somewhat unfavorable for \redsim{} as it generates a large amount of rejection events when only a small fraction of agents are infected. Consider, for instance, an agent with many neighbours of which only a few are infected. The over-approximation essentially assumes that \emph{all} neighbors are infected to sample the next event time (and, in addition, over-approximates the rate of each individual neighbor), leading to a high rejection probability. Nevertheless, the low computational costs of each rejection event overcome this. 

\vspace{-0.5cm}
\section{Conclusions\label{conclusion}}
We presented a rejection-based algorithm for the simulation of non-Markovian agent models on networks. The key advantage of the rejection-based approach is that in each simulation step it is no longer necessary to update the rates of neighboring agents. 
This greatly reduces the time complexity of each step compared to previous approaches and makes our method viable for the simulation of dynamical processes on real-world networks. 
As future work, we plan to automate the computation of the over-approximation $\widehat{\lambda}$ and investigate correlated time delays \cite{jo2019copula,masuda2018gillespie} and self-exiting point processes \cite{ogata1981lewis,dassios2013exact}.
\vspace{-0.5cm}
\subsubsection{Acknowledgements.}
We thank Guillaume St-Onge for helpful comments on non-Markovian dynamics.
This research was been partially funded by the German Research Council (DFG)
as part of the Collaborative Research Center
\say{Methods and Tools for Understanding and Controlling Privacy}.

\appendix
\section{Correctness\label{app:correctness}}

First, consider the rejection-free version of the algorithm:

\begin{enumerate}
\item Take the first event $(n,\widehat{\mu},\widehat{t}_n)$ from the event queue and update $t_{\text{global}}=\widehat{t}_n$.
\item Evaluate the true instantaneous rate  $\mu = \phi_n \big(S(n),t_{\text{global}}-T(n),M(n)\big)$ of $n$ at the current system state.
\item With probability $1-\frac{\mu}{\widehat{\mu}}$, \textbf{reject} the firing  and go to Line \texttt{5}. 
\item Randomly choose the next state $s'$ of $n$ according to the distribution \\ $\psi_n \big(S(n),t_{\text{global}}-T(n), M(n)\big)$. If $S(n)\neq s'$: set $S(n)=s'$ and $T(n)=t_{\text{global}}$.
\item Generate a new event for agent $n$ and push it to the event queue. 
\item For each neighbor $n'$ of $n$: Remove the event corresponding to $n'$ from the queue and generate a new event (taking the new state of $n$ into account).
\item Go to Line \texttt{1}. 
\end{enumerate}

Rejection events are not necessary in this version of the algorithm because all events in the queue are generated by the \say{real} rate and are therefore consistent with the current system state.
It is easy to see that the rejection-free version is a direct event-driven implementation of the \emph{Na\"ive Simulation Algorithm} which specifies the semantics in Section \ref{naive}.
The correspondence between Gillespie-approaches and even-driven simulation exploited in literature, for instance in \cite{kiss2016mathematics}.
Thus, it is sufficient to show that the rejection-free version and \redsim{} (Section \ref{algo}) are statistically equivalent. 

We do this with the following trick: We modify $\phi_n$ and $\psi_n$ into $\widehat{\phi}_n$ and $\widehat{\psi}_n$, respectively. When we simulate the rejection-free algorithm, it will admit exactly the same behavior as \redsim{}. The key to that are so-called \emph{shadow-process} \cite{cota2017optimized,grossmann2019rejection}. 
A \emph{shadow process} does not change the state of the corresponding agent but still fires with a certain rate. They are conceptually similar to self-loops in a Markov chain. In the end, we can interpret the rejection events not as rejections, but as the statistically necessary application of the shadow process.

Here, we consider the case where a constant upper-bound $c \in \mathbb{R}_{\geq 0}$ exits for all $\phi_n$. That is, $c \geq \phi_n(s,t,m)$ for all reachable $s,t,m$. The case of an time-dependent upper-bound is, however, analogous. 
Now, for each $n$, we define the shadow-process $\tilde{\phi}_n$ as 
\begin{equation*}
\widetilde{\phi}_n(s,t,m) =  c - \phi_n(s,t,m)  \;. 
\end{equation*}
Consequently, for all $n,s,t,m$:
\begin{equation*}
\widehat{\phi}_n(s,t,m) = c = \phi_n(s,t,m) + \widetilde{\phi}_n(s,t,m)
\end{equation*}
The only thing remaining is to define $\widehat{\phi}_n$ such that the shadow-process really has no influence on the system state. 
Therefore, we simply trigger a \emph{null event} (or self-loop) with the probability proportional to how much of $\widehat{\phi}_n$ is induced by the shadow-process. Formally,

\begin{equation*}
\widehat{\psi}_n(s,t,m)= 
    \begin{cases}
         p(s)=1  \text{\;\;\;\;\;\;(self-loop) with probability\;\;} \frac{\widetilde{\phi}_n}{\widehat{\phi}_n} \\
         \psi_n(s,t,m)  \text{\hspace{1.75cm} otherwise} 
    \end{cases}
    . 
\end{equation*}

Note that, firstly, the model specification with $\widehat{\phi}_n,\widehat{\psi}_n$ or  ${\phi}_n,{\psi}_n$ are equivalent, because $\widetilde{\phi}, \widetilde{\psi}$ has to actual effect on the system. Secondly, simulating the rejection-free algorithm with $\widehat{\phi}_n,\widehat{\psi}_n$ directly yields \redsim{}.
In particular, the rejections events have the same likelihood as the shadow-process being chosen in $\widehat{\psi}$. Moreover, updating the rates of all neighbors is redundant because all the rates remain at $c$. Whatever the change in $\phi_n$ is, after an event, that shadow process balances it out, such hat it actually remains constant. 

For the case that an upper-bound $c$ does not exits, we can still look at the limit case of $c \rightarrow \infty$. In particular, we truncate all rate functions at $c$ and find that, as $c$ approaches infinity, the simulated model approaches the real model.

\section{Time-Complexity\label{app:runtime}}
Next, we discuss how the runtime of \redsim{} scales with the size of the underlying contact network (and number of agents).
Assume that a binary heap is used to implement the event queue and that the graph structure is implemented using a hashmap. Each step starts by popping an element from the queue which has constant time complexity. Next, we compute $\mu$. Therefore, we have to lookup all neighbors of $n$ in the graph structure iterate over them. We also have to lookup all states and residence times. This step has linear time-complexity in the number of neighbors. More precisely, lookups in the hashmaps have constant time-complexity on average and are linear in the number of agents in the worst case. 
Computing the rejection probability has constant time complexity. In the case of a real event, we update $S$ and $T$. Again, this has constant time-complexity on average. Generating a new event does not depend on the neighborhood of an agent and has, therefore, constant time-complexity. Note that this step can still be somewhat expensive when it requires integration to sample $t_e$ but not in an asymptotic sense.
Thus, a step in the simulation is linear in the number of neighbors of the agent under consideration. 

In contrast, previous methods require that after each update, the rate of each neighbor $n'$ is re-computed. The rate of $n'$, however, depends on the whole neighborhood of $n'$. 
Hence, it is necessary to iterate over all neighbors $n''$ of every single neighbor $n'$ of $n$.

\end{document}